\documentclass[runningheads,a4paper]{llncs}

\usepackage{amssymb}
\usepackage{graphicx}

\usepackage{url}
\urldef{\mailsa}\path|{alfred.hofmann, ursula.barth, ingrid.haas, frank.holzwarth,|
\urldef{\mailsb}\path|anna.kramer, leonie.kunz, christine.reiss, nicole.sator,|
\urldef{\mailsc}\path|erika.siebert-cole, peter.strasser, lncs}@springer.com|    
\newcommand{\keywords}[1]{\par\addvspace\baselineskip
\noindent\keywordname\enspace\ignorespaces#1}

\begin{document}

\mainmatter  

\title{Internet of Cloud: Security and Privacy issues}

\titlerunning{Internet of Cloud: Security and Privacy issues}

%
%
\author{Allan Cook\inst{1}
\and Michael Robinson\inst{2}
\and Mohamed Amine Ferrag\inst{3}
\and Leandros A. Maglaras\inst{1}
\and Ying He\inst{1}
\and Kevin Jones\inst{2}
\and Helge Janicke\inst{1}
}

\authorrunning{Allan Cook et al.}

\institute{
De Montfort University, School of Computer Science and Informatics, \\ Leicester, U.K., 
\and Airbus Group Innovations, Newport, U.K., 
\and Guelma University, Department of Computer Science, Guelma, Algeria}

\toctitle{Lecture Notes in Computer Science}
\tocauthor{Authors' Instructions}
\maketitle

\begin{abstract}
The synergy between the cloud and the IoT has emerged largely due to the
cloud having attributes which directly benefit the IoT and enable its continued growth. IoT adopting Cloud services has brought new security challenges. In this book chapter, we pursue two main goals: 1) to analyse the
different components of Cloud computing and the IoT and 2) to present security and
privacy problems that these systems face. We thoroughly investigate current security
and privacy preservation solutions that exist in this area, with an eye on the Industrial
Internet of Things, discuss open issues and propose future directions.

\end{abstract}

\section{An Introduction to Cloud Technologies}

According to forecasts from Cisco Systems, by 2020 the Internet will consist of over 50 billion connected devices, including, sensors, actuators, GPS- and mobile-enabled devices, and further innovations in smart technologies, although this forecast is disputed \cite{nordrum2016popular}. New projections talk about 20 to 30 billion connected devices which is again a huge number \cite{ericsson2015ericsson}. These revolutionary devices are predicted to integrate to form hybrid networks based upon concepts such as the Internet of Things (IoT), Smart Grids, sensor networks etc., to deliver new ways of living and working.  Underpinning such operating models will be 'cloud computing' technology that enables convenient, on-demand, and scalable network access to a pool of configurable computing resources.  This remote access to high levels of processing power and storage provides a complementary platform on which to augment the low-power, low-storage characteristics of IoT devices, providing an integrated environment to provide ubiquitous capabilities to end users.

Cloud computing further offers a range of attractive benefits to organisations wishing to optimise their IT resources, such as increases in efficiency and organisational ability, reduced time to market (TTM), and a better balance between capital expenditure (capex) versus operational expenditure (opex) \cite{hausman2013cloud}.  However, to achieve such returns on investment, organisations require a clear understanding of cloud technologies to drive their strategy, and in particular, the issues surrounding security and privacy.  This chapter introduces the reader to cloud computing technologies in general, then proceeds to explain the emerging Internet of Cloud (IoC) before discussing the security and authentication issues of IoT and finally exploring the issues related to the preservation of privacy in the IoC.

\subsection{A Definition of Cloud Computing}
Cloud computing is a technological and operational model for ubiquitous, on-demand network access to a shared pool of configurable infrastructure, processing, storage and application services that can be provisioned and released for use with minimal system management effort or service provider interaction \cite{mell2011nist}.  Many of the technologies that underpin cloud computing environments are not new, as they comprise existing virtualisation, processing and storage capabilities that have existed for many years.  It is the operating model that surrounds the use of these technologies that delivers the revolutionary services,  where ownership of physical resources rests with one party, and the service users are billed for their use \cite{itpreneurs2014cloud}.

As such, it is necessary to consider the essential characteristics, service models and deployment models of cloud computing.

\subsection{Characteristics of Cloud Computing}
Cloud computing environments comprise five essential characteristics; On-demand self-service, broad network access, resource pooling, rapid elasticity, and measured service \cite{mell2011nist}.  We shall now review each of these in turn.

\begin{itemize}
\item{\bf On-demand Self-service}: In cloud environments, a consumer can request and deploy processing and storage capabilities, such as server capacity and storage space, through the use of automated provisioning services that require no negotiation with the cloud provider \cite{mell2011nist}.  This allows connected devices to remotely exploit such resources and extend their processing capabilities as necessary.

\item{\bf Broad Network Access}: The services of a cloud are made available over the network using thick or thin clients, allowing devices using different operating systems and platforms to access common capabilities \cite{mell2011nist}.

\item{\bf Resource Pooling}: The computing resources of a cloud service are pooled into a model to serve multiple consumers, with different physical and virtual resources dynamically assigned and reassigned according to demand requirements, irrespective of their geography.  The customer is typically unaware of the exact location of the provided resources, although they may be able to define high-level constraints such as country or data centre \cite{mell2011nist}.

\item{\bf Rapid Elasticity}: Elasticity is the ability of a cloud provider to scale up or down dependent upon consumer demand, allocating or freeing resources as necessary.   To the consumer, the capabilities provided often appear to be unlimited.

\item{\bf Measured Service}: In cloud models, consumers pay for the services they use, so it is necessary to monitor, control and report upon the consumption of the infrastructure.  This allows usage to be optimised, and provides a transparent understanding to both the provider and consumer \cite{mell2011nist}.
\end{itemize}

\subsection{Cloud Service Models}

There are various levels of service model available to consumers when they adopt cloud services, each with their own operating paradigm, offering software, platforms, or infrastructure as a service.  

\begin{itemize}
\item{\bf Software as a Service (SaaS)}: In this model, the consumer uses the provider's applications that run on the cloud infrastructure.  The consumer accesses these applications without any knowledge of the underlying infrastructure, and does not request or provision any associated services.  They provision and consume application resources, typically against an agreed service level agreement (SLA) that determine performance, and the cloud provider scales the infrastructure to meet its obligations \cite{mell2011nist}.
\item{\bf Platform as a Service (PaaS)}: In a PaaS environment, consumers deploy their own (or their acquired) applications, services, libraries or tools, which they control.  The cloud provider's role is to provision and maintain sufficient computing, network and storage resources to meet the agreed SLAs for the consumer-deployed elements \cite{mell2011nist}.
\item{\bf Infrastructure as a Service(IaaS)}: This service model allows consumers to provision processing, network and storage resources as necessary, onto which they can deploy whichever applications and services the require.  The consumer does not control the underlying hardware infrastructure, but can determine the technical detail of what is deployed, such as operating systems etc. \cite{mell2011nist}.
\item{\bf Cloud Deployment Models}: The provision of SaaS, PaaS or IaaS is dependent upon the cloud provider's business model.  Similarly, the scope of the cloud itself, whether private, community, public, or hybrid mix of these three, allows consumers to constrain the exposure of their information.   Irrespective of the combination of these choices however, the provider should offer the five essential characteristics of cloud services, as illustrated in Fig. 1.

\begin{figure}[h]
	\centering
	\includegraphics[width=100mm, trim=4 4 4 4,clip]{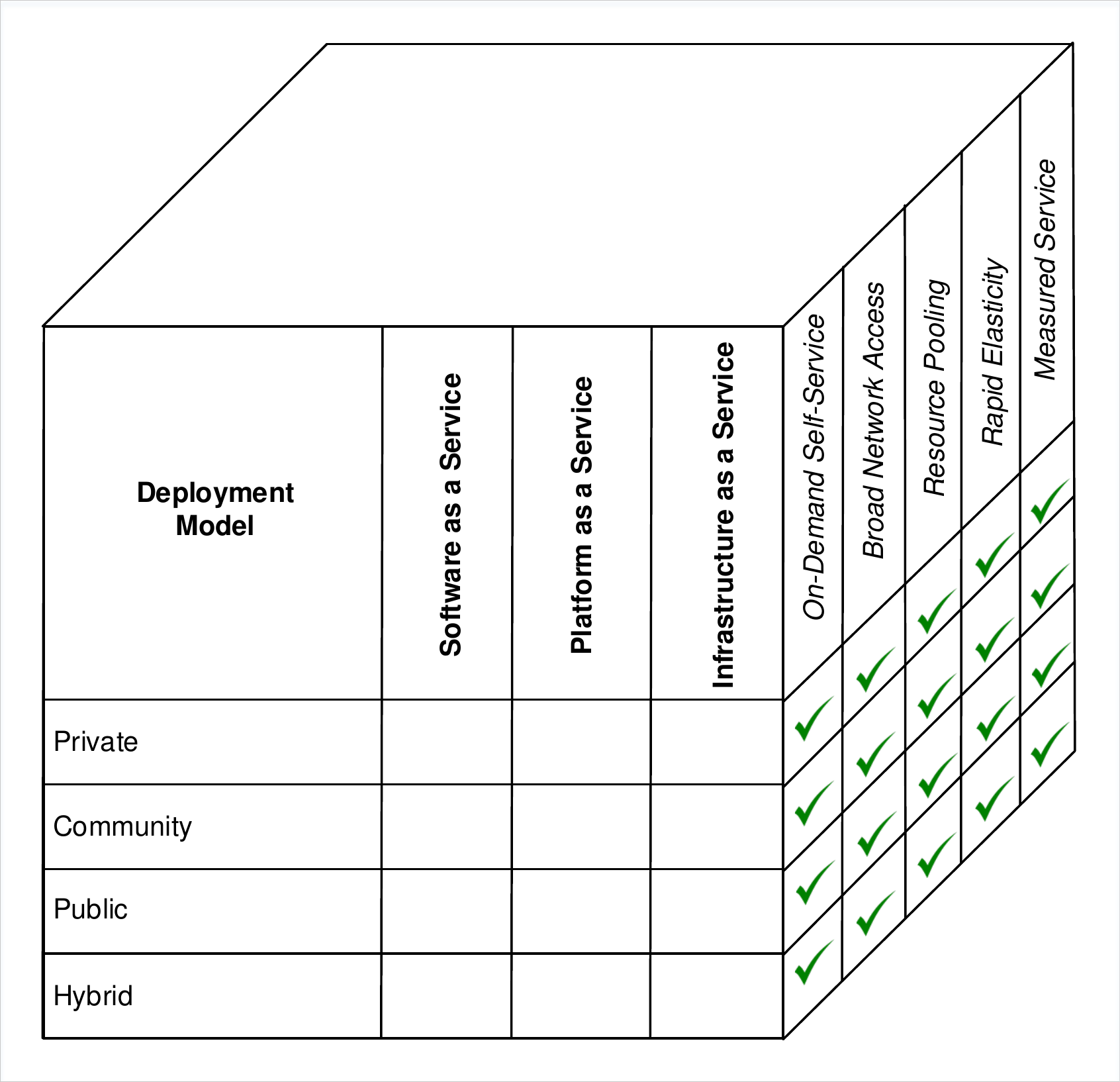}
	\caption{Cloud Deployment and Service Models Mapped to Essential Characteristics}
\end{figure}

\begin{itemize}
\item{\bf Private Cloud}: The service infrastructure is provided for exclusive use by a single organisation.  It may be owned, managed, and operated by the organisation, a third party, or some combination thereof, and may exist on or off the organisation's premises \cite{mell2011nist}.
\item{\bf Community Cloud}: The cloud is made available for use by a specific community of consumers with shared requirements.  The service may be  owned, managed, and operated by one or more of the community organisations, a third party, or some combination, and be located either on or off the premises of the community \cite{mell2011nist}.
\item{\bf Public Cloud}: The cloud infrastructure is provisioned for use by the general public.  The infrastructure is deployed on the premises of the cloud provider \cite{mell2011nist}.
\item{\bf Hybrid Cloud}: The cloud infrastructure is a mix of two or more  cloud deployment models (private, community, or public). \cite{mell2011nist}
\end{itemize}
\end{itemize}

\subsection{Enabling Technologies}
As previously discussed, cloud services are based upon a common set of underpinning enabling technologies that were developed before cloud computing emerged as a business model.  We shall now consider a key subset of these technologies in the context of their operation within a cloud.

\begin{itemize}
\item{\bf Virtualisation}: Virtualisation is the ability to deploy multiple host operating environments on one physical device.  Typically, operating systems are encapsulated within a 'virtual machine' (VM), a number of which are deployed onto a single physical server (a 'real machine').  A 'hypervisor' that abstracts the VMs from the real machine, accessing hardware components of the server as required by each VM.  Hypervisors also allow VMs to be redeployed to other real machines, permitting them to be reallocated to servers with greater or lesser processing capacity as required by the consumers demands \cite{daniels2009server}.

\item{\bf Storage}: Storage within cloud environment can be characterised as either file- or block-based services, or data management comprising record-, column- or object-based services .  These typically reside on a storage area network (SAN) that provides a persistence platform that underpins a data centre.  For file- or block-based services, the cloud ensures that sufficient capacity is provided to support the elasticity of the service, expanding or contracting as required.  Record-, column- or object-based services, however, focus on database persistence and the performance of the data used by applications.  As data expands within a large database it becomes necessary to optimise the storage based on frequency of access and location of consumers.  Data within the cloud can be easily replicated to provide temporary copies in caches etc. that improve performance, as well as reducing the impact of backup services on production data.  Similarly, where multiple data centres are used, these local caches can be optimised to focus on the datasets most frequently accessed in each location.  The underlying file system elastically supports these replicas of data, expanding and contracting as necessary to support performance SLAs \cite{grossman2009compute}.

\item{\bf Monitoring and Provisioning}: The ability of a cloud provider to automatically provision services is an key element of its offering.  Automated provisioning is typically based on a catalogue, from which consumers can select the extension or contraction of a service over which they have decided to maintain control.  The nature of the service they maintain control of is dependent upon the service model they operate within (SaaS, PaaS, IaaS).  Similarly, for the cloud provider, they require the ability to modify the execution environment in line with agreed SLAs, with or without human intervention.  The provisioning is typically managed by a service orchestration layer that interacts with the monitoring service to determine the levels of performance of cloud elements in line with SLAs, and coordinates the manipulation of the infrastructure, deploying and redeploying resources as necessary to maintain a balanced and cost-efficient use of the available architecture \cite{kirschnick2010toward}.

\item{\bf Billing}: Given the differing service and deployment models that cloud providers can offer, the billing service must be integrated with the monitoring and provisioning to ensure accurate accounting of consumption.  The billing services, in some cases, support both prepay and postpay models, requiring the billing service to decrement or accrue respectively.  The service must also only account for consumption as it occurs, and be cognisant of the elasticity of deployment and release.  As the nature of the cloud service provided to consumers may differ, the billing service must support multiple, and in many cases, complex pricing models to ensure accurate accounting \cite{elmroth2009accounting}.
\end{itemize}

Cloud computing is based on a mix of technologies brought together in differing service and deployment models to provide cost-effective utilisation of IT resources.  The ubiquity of access to these resources allows low-power, low-storage capacity IoT devices to extend their capabilities by leveraging these services on-demand.  This integration of IoT and cloud into the Internet of Cloud (IoC) provides opportunities to provide a revolution in the use and exploitation of smart devices.

\section{Internet of Things (IoT)  and Internet of Cloud (IoC)}
\label{sec:1}
The Internet of Things is a term that has rapidly risen to the forefront of the IT world, promising exciting opportunities and the ability to leverage the power of the internet to enhance the world we live in.  The concept itself is not a new one however, and it is arguable that Nikola Tesla predicted the rise of IoT back in 1926 when he stated:

\begin{quotation}
When wireless is perfectly applied the whole earth will be converted into a huge brain...and the instruments through which we shall be able to do this will be amazingly simple compared with our present telephone.~\cite{Kennedy1926}.
\end{quotation}

What Tesla had predicted was the Internet of Things (IoT), which today has been defined as the pervasive presence in the environment of a variety of things, which through wireless and wired connections and unique addressing schemes are able to interact with each other and cooperate with other things to create new applications/services and reach common goals~\cite{IERC2016}.  Put more simply, things are anything and everything around us that are internet connected and are able to send, receive or communicate information either to humans or to other things.   There are three primary methods in which things communicate.  Firstly, a sensor can communicate machine to machine (M2M).  Examples here include a sensor feeding data to an actuator which opens a door when movement is detected.  Secondly, communication can be Human to Machine (H2M), such as a sensor which can detect human voice commands.  Finally, machine to human (M2H) communication provides the delivery of useful information in an understandable form such as a display or audio announcement.  When considering the number of things in our world, and the number of potential combinations of connecting them, the only limit for thinking up valuable use cases is our own imagination.  Some well established use cases for the IoT are as follows:

\begin{itemize}
	\item \textbf{Healthcare}: The use of sensors and the internet to monitor the medical variables of human beings and perform analyses on them.  A real world example is NHS England's Diabetes Digital Coach Test Bed~\cite{NHS2015}, which trialled the use of mobile health self-management tools (wearable sensors and supporting software).  This trial leveraged the IoT to realise a number of benefits.  Firstly it enabled people with diabetes to self-manage their condition through the provision of real time data and alerts based upon the data from their sensors.  Secondly, the sensors were able to notify healthcare professionals if there was a dangerous condition that was not being corrected by the patient. Thirdly, the data from the sensors could be aggregated to provide a population-wide view of the health status of people with diabetes.
	\item \textbf{Smart Cities}: The use of connected sensors to improve city life and create an urban information system~\cite{Jin2014}.  For example, detecting the amount of waste in containers to schedule a pick-up or the use of sensors to detect city-wide availability of parking and direct drivers to an available space~\cite{Rico2013}.  This has the potential to not only save citizens frustration, but also to reduce congestion, pollution and fuel consumption.
	\item \textbf{Smart Buildings}: Both places of business and people's homes can benefit from the rise of the IoT.  Buildings consume 33\% of world energy~\cite{IERC2016}, and there is real potential for the IoT to bring this usage down.  Sensors can turn off lights when they are not needed, and appliances remotely switched off.  Heating can be optimised based upon sensors detecting occupancy and weather conditions.  Aggregations of this data can be used by energy providers to plan and optimise their operations.
	\item \textbf{Smart Transport}: The connecting of multiple transport related things.  For example, sensors in roadways and vehicles to provide a full view of traffic flow and dynamically alter traffic light sequences, speed limits, information signs or satellite navigation systems to suggest quicker routes.
	\item \textbf{Smart Industry}: Intelligent tracking of goods and components, smart factories and innovative retail concepts such as Amazon Go~\cite{Amazon2017}, which offer a checkout-less experience for customers.
\end{itemize}

\begin{figure}[ht]
\begin{center}
\includegraphics[scale=0.5]{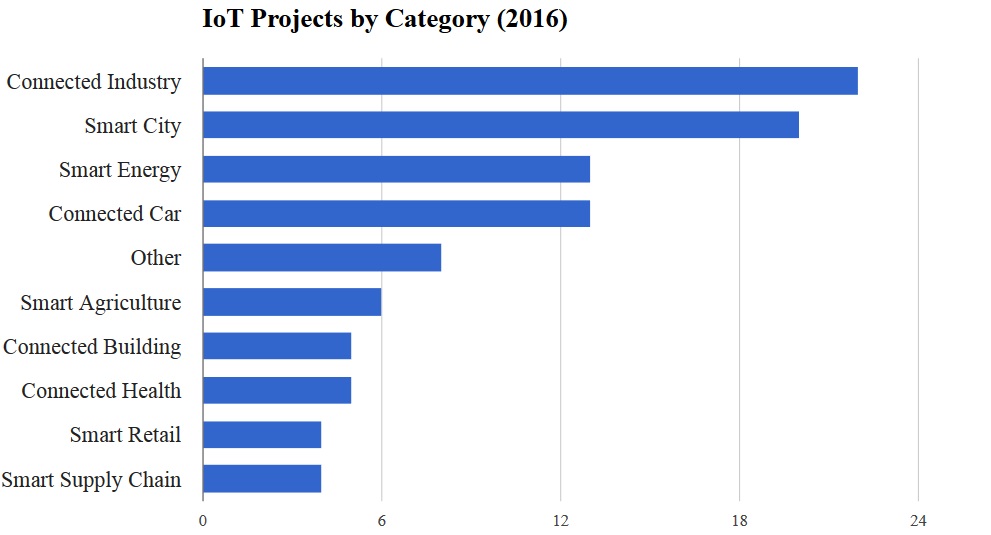}
\caption{IoT Projects by Category}
\label{fig:IoTProjects}
\end{center}
\end{figure}

A summary of IoT projects around the world ranked by application domain is provided in Figure~\ref{fig:IoTProjects}.  As the graphic shows, the most active domains for IoT (as of Q3 2016) are connected industry and smart cities.  However, all of the domains are showing an upward trend that is likely to continue into the future as new and innovative use cases are developed and the value of IoT becomes increasingly apparent to actors in each domain \cite{Bartje2016}.

\subsection{IoT Technologies}
Behind the things which make up the IoT are a number of essential core technologies.  The literature identifies five overall technologies as follows:~\cite{Lee2015}:

\begin{itemize}
	\item \textbf{Radio Frequency Identification (RFID)}: Passive RFID chips have no battery, and provide information to a reader when it is placed in proximity of the chip.  Active RFID chips can initiate communication, and output information in response to a change in its environment (e.g. changes in temperature or pressure).
	\item \textbf{Wireless Sensor Networks (WSN)}: Wireless sensor networks are defined as collections of stand-alone devices which, typically, have one or more sensors (e.g. temperature, light level), some limited processing capability and a wireless interface allowing communication with a base station~\cite{Guy2006}.
	\item \textbf{Middleware}: Middleware sits between the things and the raw data they generate to provide interoperability between things and developers who code the applications for interesting use cases.  It provides a level of abstraction, allowing developers to work with sensors without having to know the specifics of their implementation.
	\item \textbf{Cloud Computing}: The cloud provides the seemingly limitless storage and processing power necessary for IoT use cases to become a reality.
	\item \textbf{IoT Applications}: The software that provides end users with a useful product - e.g. A smartphone app through which a driver can find and reserve a free parking space.
\end{itemize}

The focus of this section is on the cloud aspect of IoT technology, in particular how cloud computing and the IoT have found what appears to be a mutually beneficial relationship and led to the term Internet of Cloud (IoC).

\subsection{Internet of Cloud (IoC)}
\label{sec:2}
The synergy between the cloud and the IoT has emerged largely due to the cloud having attributes which directly benefit the IoT and enable its continued growth.  In the IoT's infancy, things either had some local computing resources (storage, processing) to produce a useful result, or they sent their data to a mainframe which had the necessary computing resources to process the data and generate an output.  In effect, the ``brain'' as Tesla envisioned in 1926 was either highly distributed amongst the things, or it was centrally located with the things simply acting as sensors.  Both of these approaches have disadvantages.  The mainframe's weaknesses are that it is expensive to maintain and presents a central point of failure.  The highly distributed approach whereby things communicate and perform some local computation provides better resilience to failure, but increases the cost of each thing in the network.   This additional cost is both financial (the cost of equipping each thing with suitable resources and replacing failed things) and logistical (including such resources required the thing to be physically larger and consume more power).  As use cases become more advanced, and the goals more complex, the demand for more complex computation has only increased.

The IoT is not only expanding in its need for more demanding computation resources.  Gartner has predicted that the number of internet connected devices will reach 20.8 billion by 2020~\cite{Gartner2015}, suggesting that the IoT is not only expanding in computational complexity, but also in the sheer amount of data that needs to be processed.  In effect, the IoT generates big data~\cite{Rao2012}, which places the demand for smaller and cheaper things directly into competition with the demand for more computing resources.  Traditional approaches to the IoT cannot satisfy both demands - either the things become more expensive and complex, or limits on their computation resource needs are imposed.  However, the cloud presents a solution with the potential to satisfy both demands.

\subsection{Cloud as a solution}
The rise of cloud computing has provided an alternative solution, presenting the IoT with a virtually limitless source of computing power, easily accessible via the internet, with better resilience and at a lower cost than utilising a mainframe or including computing resources at the thing level.  The cloud allows IoT developers to be freed from the constraints of limited resources, and enables the use case to be realised at reduced cost.  In effect, things only require the bare minimum of hardware to perform their function (e.g. sense something, actuate something) and to communicate with the cloud.  The cloud performs all computation and communicates back the result to the things.  This pairing of cloud computing and the IoT has led to the term Internet of Cloud (IoC), and numerous literature reviews of this new paradigm are available~\cite{Botta2016,Diaz2016}.

\subsection{Sensor-Clouds}
Cloud infrastructure is not only valuable for taking on the burden of heavy computation and storage, it has also been identified as valuable in forming what are known as Sensor-Clouds~\cite{Alamri2013}.  In traditional sensor networks, the deployed sensors provide data for one purpose - to fulfil the purchaser's use case.  Unfortunately, this leads to an element of wastage, since the data being collected could be useful for other purposes but is not readily accessible by other organisations or third party developers.  For example, if a local council deployed sensors to measure traffic flow in the city centre, a third party may wish to access the sensor data to improve their satellite navigation system and direct travellers away from congested roads.  Sensor-Clouds address this scenario by making the sensor data available to multiple parties in the cloud.  In effect, they offer what could be termed sensors as a service.  This scenario brings a number of benefits.  Firstly it allows developers of IoT applications to avoid the burden of manually deploying sensors and focus upon developing interesting use cases through the use of existing sensor networks.  Secondly, sensor owners such the local council can recoup some of the cost of deployment and maintenance by charging these third parties to access the data.

\begin{figure}[ht]
\includegraphics[scale=0.8]{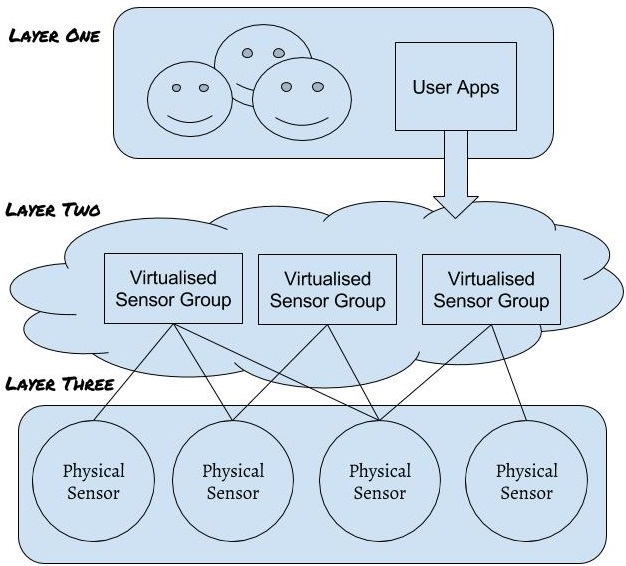}
\caption{Sensor-Cloud Layers}
\label{fig:Sensor-Cloud}
\end{figure}

A Sensor-Cloud can be visualised in three layers~\cite{Alamri2013}.  At the lowest layer, physical sensors around the world upload their data in a standardised format to the Sensor-Cloud.  This standardisation of data allows users of the service to use the data without concern over differences in areas such as protocols and formatting.  At the second layer, the Sensor-Cloud allows users to create virtualised groups of sensors for use in their applications.  These virtual sensors are based upon service templates, which are defined by the sensor owners.  At the top layer, application developers can plug these virtual sensors into their applications.  This three layer architecture for Sensor-Clouds is shown in Figure~\ref{fig:Sensor-Cloud}.

\subsection{Ongoing Challenges}
It has been noted that the cloud brings some valuable attributes to the IoT, but it is not a perfect solution which can solve all of the IoT's problems.  In fact, the use of the cloud can present some new and interesting challenges as follows~\cite{Diaz2016}:

\begin{itemize}
	\item \textbf{Security and Privacy}: Cloud security is a well documented challenge, but the pairing between cloud and the IoT presents additional concerns.  For example, when considering the sensitive nature of some use cases such as smart health, additional care must be taken to ensure that confidentially, integrity and availability of data is not violated.  Confidentiality breaches could result in personal health data being stolen, integrity breaches could be fatal if data is tampered with and a lack of availability could fail to alert to a life threatening condition.
	\item \textbf{Addressing}: If Gartner's predictions on the rapid growth of the IoT are correct, IPv4 will quickly become inadequate to address all of the things.  IPv6 has the potential to address this concern, but it is not yet widely adopted.  It has been proposed that an efficient naming and identity management system is required to dynamically assign and manage unique identity for an ever increasing number of things~\cite{khan2012}.
	\item \textbf{Latency and Bandwidth}: The cloud may provide limitless computing resources, but it cannot necessarily ensure low latency or unlimited bandwidth since this relies upon the public internet which is outside of the cloud provider's control.  This challenge has led to the rise of what is termed ``fog computing'', where computing resources are placed as close to the things as possible in order to act as an intermediary.  This intermediary can quickly service time critical processing that is latency sensitive whilst forwarding on non-time critical data for cloud processing~\cite{bonomi2012}.
	\item \textbf{Interoperability}: Due to the high number of things from multiple vendors, cloud computing alone cannot solve the issue of interoperability.  The IoT would benefit from standards which describe a common format for both handling and communicating data.
\end{itemize}

\subsection{Japan Case Study}
While the concept of Internet of Cloud and Sensor-Clouds can seem abstract, it has been implemented in some very valuable use cases.  One such use case was in the aftermath of the 2011 tsunami in Japan, which led to the second-largest nuclear emergency since Chernobyl (1986).  With a lack of reliable government information on the radiation threat, private individuals and organisations donated hundreds of Geiger counters to the affected region.  As standalone devices, the use of these counters was limited, and researchers began to investigate methods to link the devices together and make the information available to all.  The cloud provided the necessary infrastructure and agility to quickly connect each sensor, and a Sensor-Cloud of around 1000 radiation sensors was formed.  This Sensor-Cloud provided emergency services with essential information regarding radiation levels, and the same data was leveraged to produce a smart phone app for local citizens to monitor radiation levels in their area~\cite{Japan2013}.  The project, today known as Safecast~\cite{Safecast2017}, has grown beyond the initial use case and is now a global network of radiation sensors, allowing the public to access, contribute and use the sensor data through an API.

This use case highlights how valuable the Internet of Cloud can be not only for its ability to provide the necessary computing resources but also for the speed at which such resources can be provisioned - in this case rapidly providing the back end for a potentially life saving service.  As stated in the previous subsection the pairing between cloud and the IoT presents additional concerns in terms of security and privacy.  However, for these services to be adopted they must be trusted.  Therefore, we must now consider the security and privacy implications of such integration.

\section{Security and Authentication Issues}

IoT ecosystem creates a world of interconnected thing, covering a variety of application and systems, such as smart city systems, smart home systems, vehicular networks, industrial control systems as well as the interactions among them \cite{gubbi2013internet}. Cloud computing is a technology that is configured to enable access to a shared pool of resource including servers, data storage, services and application \cite{armbrust2010view}. It has become an important component of the IoT as it provides various capabilities to manage systems, servers and application and performs necessary data analytics and aggregation.  

It is undeniable that IoT provides benefits and convenience to our everyday life, however, many of the IoT components (e.g. low cost digital devices and industrial systems) are developed with little to no consideration of security \cite{jing2014security},\cite{khorshed2012survey}, \cite{samarati2016cloud}. A successful exploit can propagate within the IoT ecosystem that could render the loss of sensitive information, interruption of the business functionalities, and the damage to critical infrastructure. 

We have already seen the security concerns of the cloud services.  These include but are not limited to malware injection (malicious code injected into cloud and run as SaaS), vulnerable application programming interfaces (API), abuse of data and cloud service, insider threats and the newly emerging Man In Cloud Attack \cite{jensen2009technical}. Cloud involves both service providers and consumers; therefore, cloud security is a shared responsibility between the two. 

Security issues are still yet to be solved for IoT and Cloud respectively. IoT adopting Cloud services could complicate this situation and raise more security concerns. This section focuses on the security issues on IoT adopting Cloud services and makes recommendations to address those issues. 

\subsection{Data sharing/management issues of IoT Cloud }

Within a cloud context, no matter public, private or hybrid, data security management involves secure data storage, secure data transmission and secure access to the data. During transmission, the Transport Layer Security (TLS) cryptography is widely used to prevent against threats. During processing, the cloud service provided applies isolation between different consumers. The isolation \cite{wei2014security} is applied at different levels such as operations system, virtual machine or hardware. The secure access to data sometimes depends on the isolation level. It may sometimes separate completely from other resources or may have shared infrastructures and softwares that rely on access control policies. Within an IoT context, one of the benefits is open data sharing among different things. If the data is isolated as is currently offered in the cloud services, open wide data aggregation and analytics would become impossible. A balance needs to be found between data protection and sharing.

There is existing work such as Information Flow Control (IFC)\cite{bacon2014information,pasquier2015clouds} defining and managing the requirements for data isolation and sharing. People can specify to what extent they want to share or protect their data. Other work is related to data encryption, encrypting the things before uploading to the cloud. This would limit the users access to the data, which again affects the IoT's data sharing and analytical capability. There are some solutions to analyse encrypted data. However, this approach is not mature to be applied in practice at this stage \cite{naehrig2011can,hrestak2014homomorphic}.

\subsection{Access Control and Identity Management}

Within a cloud context, access control rules are strictly enforced. The service providers use authentication and authorisation services to grant privileges to the consumers to access storage or files. Within an IoT context, there are interactions between different devices that are owned by different people. Access control is usually leveraged through device identity management and configuration \cite{mahalle2010identity}. Existing identity management includes identity encoding and encryption \cite{vermesan2011internet}.

When IoT uses Cloud services, access control involves the interactions among the applications and cloud resources. Policies and mechanism need to be flexibly defined to accommodate the needs of both and resolve the conflicts of different parties. There is existing work on grouping the IoT devices to enable common policies \cite{savola2013metrics}. However, cares need to be taken to ensure the flexibly defined policies do not introduce vulnerabilities to the system.

\subsection{Complexity (Scale of the IoT)}

One of the benefits of Cloud service adoption is the reduction of cost through elastically resource scaling mechanisms. The increase of IoT devices, data volume, and variety has become a burden for the Cloud. The failure to coordinate and scale the "things" will impact the availability of the data. Security mechanism will bring extra burden that can impact the performance of IoT Cloud \cite{wei2014security,rittinghouse2016cloud}.

Logging is an important aspect of security as it provides a view of the current state of the system. Logging within the Cloud is centralised and it is an aggregation of the logs from different components such as software applications and operation \cite{singh2016twenty}. IoT logging tends to decentralise it among different components. There are some existing work on logging centralisation (e.g. design analytics tools to collect and correlate decentralised logs from "things" \cite{rabkin2010chukwa}) and decentralisation (e.g. enable logging to capture information in a data-centric manner \cite{rabkin2013making}). A balance needs to be found between logging centralisation and decentralisation.
 
\subsection{Protection of different parties}

IoT Cloud raise security concerns to both service providers and consumers. Cloud service providers used to apply access control to protect data and resources. Now, the "things" can directly interact with the providers. Attacks can be easily launched by compromised "things". We have already seen some real world exploits of smart home applications, that are designed with poor security considerations \cite{robles2010review}.

From the consumers' perspective, "things" needs to be validated before it can be connected. If the "things" are determined to be compromised or malicious, alerts will be sent either in a human readable or machine-readable format. The determination can be based on reputation, trustworthy network node evaluation \cite{yan2014survey,nitti2014trustworthiness} and so on.

\subsection{Compliance and legal issues}

Cloud demonstrated compliance using contract through service-level agreement (SLA). A typical method to assess compliance is through auditing. Within the area of Cloud, Massonet has proposed a framework that can generate auditing logs demonstrating that they are compliant with the related policies/regulations \cite{massonet2011monitoring}. There are also frameworks designed in the area of IoT to demonstrate compliance using auditing logs.  

IoT tends to be decentralised in isolated physical locations. The centralisation of cloud allows the data to flow across geographic boundaries, which has raised legal and law concerns of the data across national borders. There are some existing work on constrain data flow geographically by applying legal and management principles to the data \cite{singh2014regional}. Again this will have an negative impact on data sharing capability of IoT and Cloud.

\subsection{Cloud decentralisation}

An emerging trend is the Cloud decentralisation in order to accommodate the IoT and big data analytics. Typical method is decentralised computing such as Fog computing \cite{bonomi2012fog} and grid computing \cite{foster2008cloud}. Cloud decentralisation helps reduce the typical Cloud attacks such as denial of service (DoS) attack; it also raises new security concerns. Instead of targeting on the Cloud services, the attacks are directed to individual service providers and consumers. There is on-going research in securing the decentralised Cloud through coordinating the communication of things to things, things to clouds and clouds to clouds \cite{singh2014policy,singh2016twenty}. Finally cloud decentralization can provide more flexible management. 

Following the trend of decentralized cloud deployment, security mechanisms that must be developed for the IoC may also be decentralized. This deployment can have multiple advantages and in this context Intrusion Detection Systems for the IoC are analyzed on Section \ref{sec_IDS}.

\subsection{Intrusion Detection Systems}\label{sec_IDS}

Intrusion detection systems (IDS) can be classified into centralized intrusion detection systems (CIDS) and distributed intrusion detection systems (DIDS) by the way in which their components are distributed. In a CIDS the analysis of the data is performed  in  some  fixed  locations  without  considering  the  number  of  hosts  being monitored \cite{kumar1995classification}, while a DIDS is composed of several IDS over large networks whose data analysis  is performed  in  a  number  of  locations proportional  to  the  number  of  hosts. There are numerous advantages of a DIDS compared to a CIDS. A DIDS  is highly  scalable  and  can provide  gradual degradation  of  service, easy extensibility and scalability \cite{crosbie1995active}. 

Novel intrusion detection Systems (IDS) must be implemented that need to be efficient both in terms of accuracy, complexity and communication overhead, false alarm rate and time among others. IDSs that have been developed for other systems, e.g. Industrial Control Systems \cite{maglaras2016combining,cruz2016cybersecurity}, wireless sensor networks \cite{butun2014survey}, or cloud environments \cite{modi2013survey} can be used as a basis for developing new detection systems for the IoC area. Adaptivity of the IDS on changes in the network topology, which will be a frequent situation for an IoC, is an aspect that needs to be addressed when new IDS are going to be designed \cite{stewart2017eai}.

\section{Privacy preserving schemes for IoC}
\begin{figure}[h]
 \centering
 \includegraphics[width=1\linewidth]{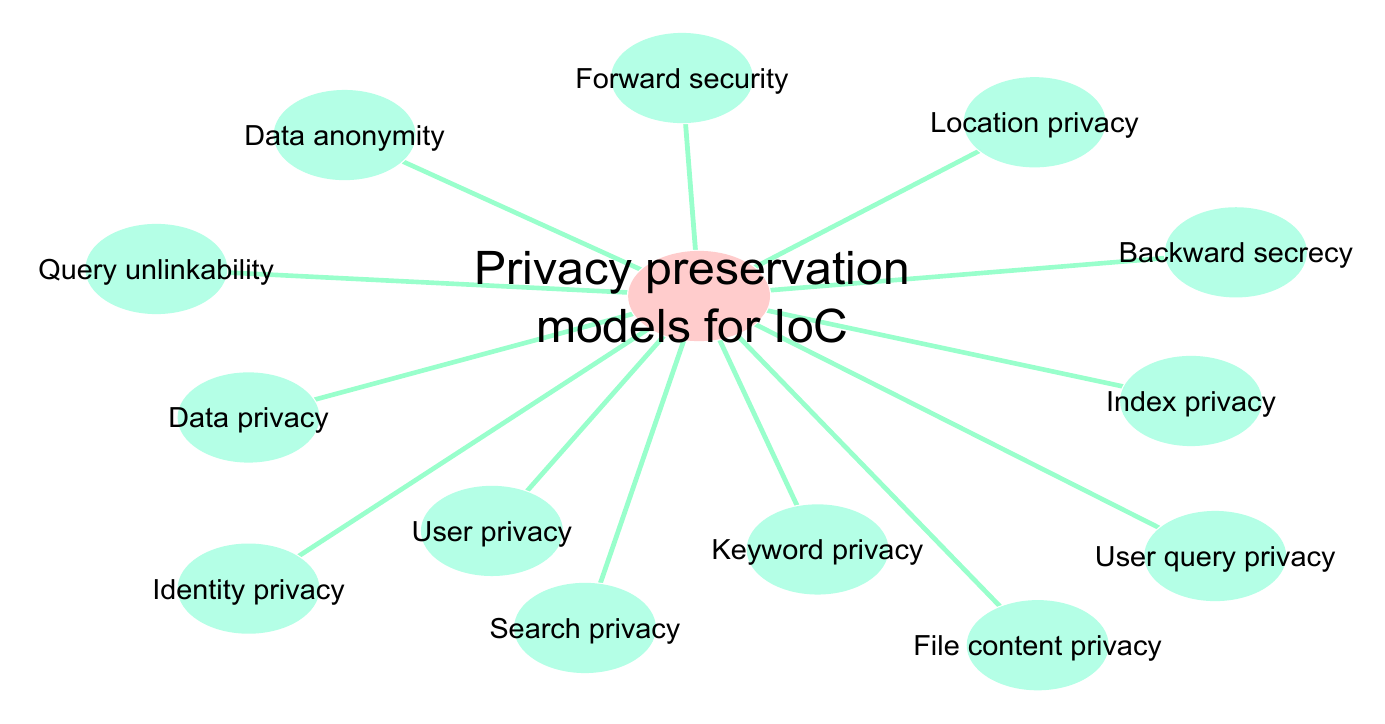}
 \caption{Classification of privacy preservation models for IoC}
 \label{fig:Fig4a}
 \end{figure}

In this subsection, we review the privacy preserving schemes for IoC. Based on the classification of authentication and privacy preserving schemes in our three recent surveys \cite{F19,F20,F21}, the privacy preserving schemes for IoC can be classified according to networks models and privacy models. The summary of privacy preserving schemes for IoC are published in 2014, 2015, and 2016, as presented in Tab. \ref{tab:4a}, Tab. \ref{tab:4b}, and Tab. \ref{tab:4c}, respectively. In addition, Fig. \ref{fig:Fig4a} shows the classification of privacy preservation models for IoC.

\begin{table}
\begin{center}
\scalebox{0.73}{
\begin{tabular}{||p{0.4in}|p{1in}|p{0.9in}|p{0.9in}|p{1.2in}|p{1.5in}||} \hline 
\textbf{Scheme} & \textbf{System Model} & \textbf{Privacy model} & \textbf{Goals} & \textbf{Main phases } & \textbf{Performances (+) and limitations (-)} \\ \hline 
Cao et al. \cite{F1} & A cloud data hosting service involving three different entities, namely, the data owner, the data user, and the cloud server &  Data privacy;\newline  Index privacy;\newline  Keyword privacy; & Achieving the multi-keyword ranked search with privacy-preserving & - Setup;\newline - BuildIndex;\newline - Trapdoor;\newline - Query; & + Efficient in term of the time cost of building index;\newline + Efficient in term of the time cost of query;\newline + Resistance to the known ciphertext model and known background model;\newline - No consideration for checking the integrity of the rank order. \\ \hline 
Worku et al. \cite{F2} & Three different entities, including, cloud server, user, and third party auditor & - User privacy & Achieving public verifiability, storage correctness, batch auditing, blockless verification, and privacy preserving &  KeyGen;\newline  SigGen;\newline  ProofGen;\newline  VerifyProof; & + Efficient in term auditor and server computation overhead compared to the scheme in \cite{F22};\newline + Secure in the random oracle model;\newline - No threat model presented. \\ \hline 
Wang et al. \cite{F3}  & Three different entities, including, data owner, cloud server, and user & - File content privacy;\newline - Index privacy;\newline - User query privacy; & Support multi-keyword fuzzy search & - KeyGen;\newline - Index Enc;\newline - Query Enc;\newline - BuildIndex;\newline - Trapdoor;\newline - Search; & + Efficient in term of the Bloom filter generation time for a single file;\newline + Multi-keyword fuzzy search;\newline + Resistance to the known ciphertext model and known background model;\newline - No comparison with related schemes. \\ \hline 
Wang et al. \cite{F4} & Three parties: the cloud server, a group of users and a public verifier & - Identity privacy; & Achieving following properties: (1) Public Auditing, (2) Correctness, (3) Unforgeability, and (4) Identity Privacy. & - KeyGen;\newline - RingSign;\newline - RingVerify; & + Efficient in terms of signature generation and communication cost;\newline + Efficient in term of auditing time;\newline + Efficient in terms of privacy and batch auditing with incorrect proofs;\newline - Traceability is not considered. \\ \hline 
Yuan and Yu \cite{F5} & Three major parties: a trusted authority (TA), the participating parties (data owner) and the cloud servers (or cloud) & - Multi-party privacy-preserving; & Protecting each participant's private dataset and intermediate results generated during the back-propagation neural network learning process & - Privacy Preserving Multi-Party Neural Network\newline Learning;\newline - Secure Scalar Product and Addition with Cloud; & + Efficient in terms of collaborative learning and communication cost compared to both schemes in \cite{F24} and \cite{F25};\newline + Support the multi-party scenario;\newline + Scalable, efficient and secure;\newline - Does not allow multiparty collaborative learning without the help of TA. \\ \hline 
Sun et al. \cite{F6} & Three entities: the data owner, the data user, and the cloud server & - Search privacy;\newline - Index confidentiality; - Query confidentiality;\newline - Query unlinkability;\newline - Keyword privacy; & Achieving high efficiency and functionality (such as expressive/usable queries) & - Setup;\newline - GenIndex;\newline - GenQuery;\newline - SimEvaluation; & + Efficient in term of time cost for generating encrypted query;\newline + Help users ensure the authenticity of the returned search results in the multi-keyword ranked encrypted text search scenario;\newline - Traceability is not considered. \\ \hline 
Dong et al. \cite{F7} & Four parties in a network: the data owner, the data consumer, the cloud server, and the private key generator & - Backward secrecy;\newline - User privacy & Achieving fine-grained access control & - System initialization;\newline - Encryption;\newline - Key generation and distribution;\newline - Decryption; & + Efficient in terms of computation complexity, communication cost, and cost of revocation operation;\newline + Fully collusion secure;\newline + User access privilege confidentiality;\newline - Adversary's model is limited. \\ \hline 
\end{tabular}
 }
\end{center}
\caption{Summary of privacy preserving schemes for IoC (Published in 2014)}
 \label{tab:4a}
 \end{table}

Cao et al. \cite{F1} defined and solved the problem of multi-keyword ranked search over encrypted cloud data while preserving strict systemwise privacy in the cloud-computing paradigm. Specifically, the authors proposed a preserving scheme, called MRSE, using the secure inner product computation. The MRSE scheme is efficient in terms of the time cost of building index and the time cost of query. Worku et al. \cite{F2} proposed a privacy-preserving public auditing protocol in order to provide the integrity assurance efficiently with strong evidence that unfaithful server cannot pass the verification process unless it indeed keeps the correct data intact. Wang et al. \cite{F3} proposed a brand new idea for achieving multi-keyword (conjunctive keywords) fuzzy search. Different from existing multi-keyword search schemes, the scheme \cite{F3} eliminates the requirement of a predefined keyword dictionary. Based on locality-sensitive hashing and Bloom filters, the scheme \cite{F3} is efficient in term of the Bloom filter generation time for a single file. Wang et al. \cite{F4} a privacy-preserving public auditing mechanism, called Oruta, to verify the integrity of shared data without retrieving the entire data. Oruta uses ring signatures \cite{F23} to construct homomorphic authenticators. In addition, Oruta can achieving following properties: (1) Public Auditing, (2) Correctness, (3) Unforgeability, and (4) Identity Privacy. Yuan and Yu \cite{F5} proposed the first secure and practical multi-party the back-propagation neural network learning scheme over arbitrarily partitioned data. Based on two phases, namely, 1) privacy preserving multi-party neural network learning, and 2) secure scalar product and addition with Cloud, the scheme \cite{F5} can support the multi-party scenario and efficient in terms of collaborative learning and communication cost compared to both schemes in \cite{F24} and \cite{F25}. Sun et al. \cite{F6} proposed an idea to build the search index based on the vector space model and adopt the cosine similarity measure in the Cloud supporting similarity-based ranking. Based on the vector space model, the scheme \cite{F6} is efficient in term of time cost for generating encrypted query. Dong et al. \cite{F7} considered four parties in a network, namely, the data owner, the data consumer, the cloud server, and the private key generator. Then, the authors \cite{F7} proposed an idea that the cloud can learn nothing about a user's privacy or access structure, as such the scheme is fully collusion resistant. 

\begin{table}
\begin{center}
\scalebox{0.73}{
   \begin{tabular}{||p{0.4in}|p{1in}|p{0.9in}|p{0.9in}|p{1.2in}|p{1.5in}||} \hline 
\textbf{Scheme} & \textbf{System Model} & \textbf{Privacy model} & \textbf{Goals} & \textbf{Main phases } & \textbf{Performances (+) and limitations (-)} \\ \hline 
Zhou et al. \cite{F8} & Cloud-assisted wireless body area networks & - Identity privacy;\newline - Location privacy; & Detecting two attacks, namely, time-based mobile attack and location-based mobile attack & - Pairwise key establishment;\newline - Group key agreement; & + Efficient in terms of storage, computation, and communication overhead compared to the scheme in \cite{F26};\newline - Traceability is not considered;\newline - No consideration for the patients' selfishness. \\ \hline 
Zhou et al. \cite{F9}\newline  & Three components: body area networks(BANs), wireless transmission networks and the healthcare providers equipped with their own cloud servers & - Identity privacy;\newline - Data confidentiality; & Achieving data confidentiality and identity privacy with high efficiency & - Setup;\newline - Key Extract;\newline - Sign;\newline - Verify; & + Efficient in terms of computational overhead, communication overhead, and storage overhead compared to the scheme in \cite{F27};\newline - Location privacy is not considered; \\ \hline 
Liu et al. \cite{F10}\newline  & Three main network entities: users, a cloud server, and a trusted third party & - Data anonymity;\newline - User privacy;\newline - Forward security; & Achieving authentication and authorization without compromising a user's private information & - Ideal data accessing functionality;\newline - Ideal authority sharing functionality; & + Considers the data anonymity;\newline - Need an evolution in terms of computational overhead, communication overhead, and storage overhead;\newline - Traceability is not considered;\newline - No comparison with related schemes. \\ \hline 
\end{tabular}
 }
\end{center}
 \label{tab:4b}
\caption{Summary of privacy preserving schemes for IoC (Published in 2015)}
 \end{table}
 
  To resilient to both time-based and location-based mobile attacks, Zhou et al. \cite{F8} proposed a secure and privacy-preserving key management scheme for cloud-assisted wireless body area networks. Based on the body's symmetric structure with the underlying Blom's symmetric key establishment mechanism, the scheme \cite{F8} is efficient in terms of storage, computation, and communication overhead compared to the scheme in \cite{F26}, but the traceability is not considered. Zhou et al. \cite{F9} proposed a patient self-controllable and multilevel privacy-preserving cooperative authentication scheme, called PSMPA, to realize three levels of security and privacy requirement in distributed m-healthcare cloud computing system which mainly consists of the following five algorithms: Setup, Key Extraction, Sign, Verify and Transcript Simulation Generation. Based on an attribute based designated verifier signature scheme, PSMPA is efficient in terms of computational overhead, communication overhead, and storage overhead compared to the scheme in \cite{F27}, but the location privacy is not considered. Therefore, Liu et al. \cite{F10} proposed a shared authority based privacy-preserving authentication protocol, named, SAPA, for the cloud data storage, which realizes authentication and authorization without compromising a user's private information. SAPA protocol \cite{F10} applied ciphertext-policy attribute based access control to realize that a user can reliably access its own data fields. In addition, SAPA protocol \cite{F10} adopt the proxy re-encryption to provide temp authorized data sharing among multiple users.
  
 \begin{table}
\begin{center}
\scalebox{0.73}{
  \begin{tabular}{||p{0.4in}|p{1in}|p{0.9in}|p{0.9in}|p{1.2in}|p{1.5in}||} \hline 
 \textbf{Scheme} & \textbf{System Model} & \textbf{Privacy model} & \textbf{Goals} & \textbf{Main phases } & \textbf{Performances (+) and limitations (-)} \\ \hline 
 Xia et al. \cite{F11} & Four different types of entities: the image owner, image user, cloud server and watermark certification authority & - Data privacy\newline  & Protecting the privacy of image data in content-based image retrieval outsourcing applications against a curious cloud server and the dishonest query users & - KeyGen;\newline - IndexGen;\newline - ImgEnc; & + Privacy of image content;\newline + Privacy of image features;\newline + Privacy of trapdoors;\newline + Leakage of similarity information;\newline + Efficient in term of time consumption of the trapdoor generation\newline - The proposed watermarking method cannot\newline be regarded as a very robust one. \\ \hline 
 Xia et al. \cite{F12} & Three different types of entities: the image owner, image user and cloud server & - Image privacy & The plaintext data needs to be kept unknown to the cloud server & - The generation of unencrypted index;\newline - The index encryption; & + The privacy of the image;\newline + The privacy of the image features;\newline + The privacy of the trapdoors;\newline + The leakage of the similarity information;\newline + Efficient in four terms, namely, 1) time consumption of the index construction, 2) time consumption of the trapdoor generation, 3) time consumption of the search operation, and 4) storage consumption of the index;\newline - The feature extraction in encrypted image. \\ \hline 
 Pasupuleti et al. \cite{F13} & Consisting of three main entities, including, data owner, cloud service provider, and authorized users & - Index Privacy\newline - Data privacy\newline  & Proposing an efficient and secure privacy-preserving\newline approach with following goals: privacy preserving, index privacy, and data integrity & - Key generation;\newline - Index creation;\newline - Privacy-preserving;\newline - Trapdoor generation;\newline - Ranked keyword search;\newline - Data decryption; & + Detect the modifications or deletions of data and maintain the consistency of data;\newline + Efficient in terms of computation cost and communication cost;\newline - Dynamic data updates; \\ \hline 
 Xia et al. \cite{F14} & Consisting of three main entities, including, data owner, data user and cloud server & - Index confidentiality;\newline - Query confidentiality;\newline - Trapdoor unlinkability;\newline - Keyword privacy; & Supporting multi-keyword ranked search and dynamic operation on the document collection & - Index Construction;\newline - Search Process; & + The search precision on different privacy level;\newline + The efficiency of index construction, trapdoor generation, search, and update;\newline - The users keep the same secure key for trapdoor generation;\newline - Location privacy and identity privacy are not considered. \\ \hline 
 Song et al. \cite{F15} & Consisting of three main entities, including, data owner, authorized data user, and cloud server & - Data privacy;\newline - Query privacy; & Providing the full-text retrieval services with privacy-preserving & - Document processing;\newline - Index structure and maintenance mechanism;\newline - Full-text retrieval algorithm over encrypted data; & + The index space cost;\newline + Time cost for inserting a new document;\newline + Query efficiency with different number of documents;\newline + Query precision with different number of documents;\newline - Traceability is not considered; \\ \hline 
 Zhu et al. \cite{F16} & Four parts: trusted authority, location based services (LBS) provider, LBS user, and cloud server & - Location privacy & Providing privacy-preserving LBS data and user's location information with accurate LBS for users. & - System initialization;\newline - Cloud server data creation;\newline - Privacy-preserving location based services; & + The user query location is privacy-preserving in the proposed EPQ scheme;\newline + The proposed EPQ scheme can achieve confidential LBS data; \newline + The authentication of the LBS query request and response are achieved in the proposed EPQ scheme;\newline + Efficient in term of computation complexity compared with the FINE scheme \cite{F28};\newline - Traceability is not considered; \\ \hline 
 Lyu et al. \cite{F17} & Four parts: social network, data owner, members, and cloud storage server & - Data privacy & Providing the fine-grained access control & - System initiation;\newline - Privacy-preserving data construction;\newline - Interested data acquisition;\newline - Dynamic attribute management; & + Data confidentiality;\newline + Fine-grained access control;\newline + Collusion attacks resistant;\newline + Efficient in terms of communication overhead and computation cost compared to the scheme in \cite{F29};\newline - Location privacy and identity privacy are not considered. \\ \hline 
 Yu et al. \cite{F18} & Four parts, key distribution center, cloud user, cloud server and third party auditor & - Data privacy & Formalize the security model of zero knowledge privacy against the third party auditor & - Setup;\newline - Extract;\newline - TagGen; & + Perfect data privacy preserving;\newline + Efficient in terms of costs regarding computation, communication and storage;\newline - No comparison with related schemes. \\ \hline 
 \end{tabular}
  }
	\end{center}
  \label{tab:4c}
	\caption{Summary of privacy preserving schemes for IoC (Published in 2016)}
  \end{table}
 Xia et al. \cite{F11} proposed an idea to protect the privacy of image data in content-based Image retrieval outsourcing applications against a curious cloud server and the dishonest query users. Specifically, the idea is the first work that proposes a searchable encryption scheme, considering the dishonest query users who may distribute the retrieved images to those who are unauthorized. Similarly to the scheme \cite{F11}, Xia et al. \cite{F12} proposed an idea which the secure k-nearest neighbor algorithm is employed to protect the feature vectors in order to enable the cloud server to rank the search results very efficiently without the additional communication burdens. Based on the pre-filter tables, the scheme \cite{F12} is efficient in four terms, namely, 1) time consumption of the index construction, 2) time consumption of the trapdoor generation, 3) time consumption of the search operation, and 4) storage consumption of the index. Pasupuleti et al. \cite{F13} proposed an efficient and secure privacy-preserving approach using the probabilistic public key encryption technique to reduce computational overhead on owners while encryption and decryption process without leaking any information about the plaintext. Based on the idea of integrity verification, the scheme \cite{F13} can detect the modifications or deletions of data and maintain the consistency of data, also is efficient in terms of computation cost and communication cost. Therefore, Xia et al. \cite{F14} proposed two secure search schemes, namely, 1) the basic dynamic multi-keyword ranked search scheme in the known ciphertext model, and 2) the enhanced dynamic multi-keyword ranked search scheme in the known background model. Based on the searchable encryption scheme, the idea in \cite{F14} can supports both the accurate multi-keyword ranked search and flexible dynamic operation on document collection. Song et al. \cite{F15} defined and solved the problem of full-text retrieval over encrypted cloud data in the cloud computing paradigm. Based on a hierarchical Bloom filter tree index, the scheme \cite{F15} can protect user query privacy. Zhu et al. \cite{F16} proposed a privacy-preserving location based services query scheme in outsourced cloud, called EPQ, for smart phone. EPQ scheme can achieve confidential location-based services (LBS) data and is efficient in term of computation complexity compared with the FINE scheme \cite{F28}.
 
 Lyu et al. \cite{F17} design an efficient and secure data sharing scheme, named DASS. Based on multi-attribute granularity for social applications, DASS can support searchable encryption over data, and is efficient in terms of communication overhead and computation cost compared to the scheme in \cite{F29}, but location privacy and identity privacy are not considered. Yu et al. \cite{F18} investigated a new primitive called identity-based remote data integrity checking for secure cloud storage. In addition, the scheme \cite{F18} showed that it achieves soundness and perfect data privacy. In a recent work, Ferrag and Ahmim in \cite{F30} proposed an efficient secure routing scheme based on searchable encryption with vehicle proxy re-encryption, called ESSPR, for achieving privacy preservation of message in vehicular peer-to-peer social network.

\section{Summary}

The synergy between the cloud and the IoT has emerged largely due to the cloud having attributes which directly benefit the IoT and enable its continued growth. IoT adopting Cloud services has brought new security challenges. We have identified key security issues in data management, access control and identity management, complexity and scale, the protections of different parties, compliance and legal issues, as well as the emerging Cloud decentralisation trend. There is existing work addressing these issues, however future work should primarily focus on the balance between centralisation and decentralisation, data security and sharing as well as associated policy issues. 

Regarding privacy preservation, it is not a problem that can be treated in isolation for a system, but interdependencies among different users and platforms must be also analyzed. Also the combination of privacy metrics can help improve the level of privacy by combining the positive aspects of different methods while keeping the total cost, in terms of storage, computation and delay, relatively low.

\bibliographystyle{UNSRT}
\bibliography{Cloud}

\end{document}